\documentclass[fleqn,twoside]{article}
\usepackage{espcrc2}

\usepackage{psfig}
\usepackage[figuresright]{rotating}

\newcommand{\lesssim}{\mathbin{\lower 3pt\hbox
   {$\rlap{\raise 5pt\hbox{$\char'074$}}\mathchar"7218$}}} 
\newcommand{\gtrsim}{\mathbin{\lower 3pt\hbox
   {$\rlap{\raise 5pt\hbox{$\char'076$}}\mathchar"7218$}}} 

\title{Strong-Field Tests of Gravity Using Pulsars and
Black Holes}

\author{M. Kramer\address{University of Manchester, Jodrell Bank Observatory, 
                          Jodrell Bank, UK},
\thanks{mkramer@jb.man.ac.uk},
        D. C. Backer\address{Department of Astronomy, Univsersity of
                     California at Berkeley, Berkeley, CA, USA}
       \thanks{dbacker@astro.berkeley.edu},
        J. M. Cordes\address{Department of Astronomy, Cornell University, 
                       Ithaca, NY, USA}
\thanks{cordes@astro.cornell.edu},
        T. J. W. Lazio\address[NRL]{Naval Research Laboratory, 
                     Washington, DC, USA}
\thanks{Joseph.Lazio@nrl.navy.mil}
\thanks{Basic research in radio astronomy at the NRL is supported
by the Office of Naval Research.},
        B. W. Stappers\address{ASTRON, Dwingeloo, The Netherlands}
\thanks{stappers@astron.nl}
S. Johnston\address{University of Sydney, NSW 2006, Australia}
\thanks{simonj@physics.usyd.edu.au}
}

\begin{document}

\begin{abstract}
The sensitivity of the SKA enables a number of tests of theories of
gravity.  A Galactic Census of pulsars will discover most of the
active pulsars in the Galaxy beamed toward us.  In this census will
almost certainly be pulsar-black hole binaries as well as pulsars
orbiting the super-massive black hole in the Galactic centre. These
systems are unique in their capability to probe the ultra-strong field
limit of relativistic gravity. These measurements can be used to test
the Cosmic Censorship Conjecture and the No-Hair theorem.

The large number of millisecond pulsars discovered with the SKA will
also provide a dense array of precision clocks on the sky. These
clocks will act as the multiple arms of a huge gravitational wave
detector, which can be used to detect and measure the stochastic
cosmological gravitational wave background that is expected from a
number of sources.
\end{abstract}

\maketitle


\section{Was Einstein Right?}
\label{einstein}

In astrophysical experiments we are passive observers who must derive
all information simply from photons (or particles or gravitons) received,
in contrast to procedures in terrestrial laboratories where
experimental set-up can be modified and environment can be
controlled. As a result, terrestrial experiments are typically more
precise and, most importantly, reproducible in any other laboratory on
Earth. However, when probing the limits of our understanding of
gravitational physics, we are interested in extreme conditions that
are not encountered on Earth. In some cases, it is possible to perform
the experiment from a space-based satellite observatory. Indeed, solar
system tests provide a number of very stringent tests of Einstein's
theory of general relativity (GR) (see \cite{wil01}), and to date GR
has passed all observational tests with flying colours.

Despite the success of GR, the fundamental question remains as to
whether Einstein has the last word in our understanding of gravity or
not --- a question that was also included as one of eleven questions
raised in {\em ``Connecting Quarks with the Cosmos: Eleven Science
Questions for the New Century''} \cite{bpa03}. The likely answer to
this question is that this is not the case, as physicists attempt to
formulate a theory of quantum gravity. Quantum gravity would fuse the
classical world of gravitation, currently best described by GR, with
the intricacies of quantum mechanics. Quantum gravity would therefore
account for all of the known interactions and particles of the
physical world. Determining as to whether the as yet accurate theory of 
GR describes the gravitational interaction of the
macroscopic world correctly, would either justify the current
approaches to use GR as the basis of quantum gravity or would
imply that other alternative lines of investigations have to be
followed during this enormous task.

There is a large parameter space that is not yet explored by the
current experimental tests of GR. In particular, solar-system
experiments made or proposed for the future are all made in the
weak-field regime and hence will never be able to provide tests in the
strong-field limit.  Other tests involving the observations of X-ray
line redshifts from neutron stars may explore some parts of this
parameter space, but the interpretation of these results depends to
some extent on the unknown equation-of-state \cite{dp03}. Similarly,
whilst it is in principle possible to utilize low-energy X-ray
spectra or high frequency quasi-periodic
oscillations to study relativistic effects around
a spinning and accreting BH (e.g.~\cite{ml98}), the arguments
still require certain plausible assumptions (e.g., the accretion disk
reaches all the way down to the last stable orbit, etc.), and the
precision with which the spin is determined is rather poor, even in
systems where all the input parameters are comparably well known. In
contrast, pulsars provide an accurate clock attached to a point-mass
which can, in a binary orbit, allow us to perform high precision tests
of gravitational theories.  Hence, binary pulsars are and will remain
the only way to test the predictions made by GR or competing theories
of relativistic gravity in the strong-field limit. This is possible
with a precision that is otherwise only achieved in terrestrial
laboratories.

\begin{figure*}
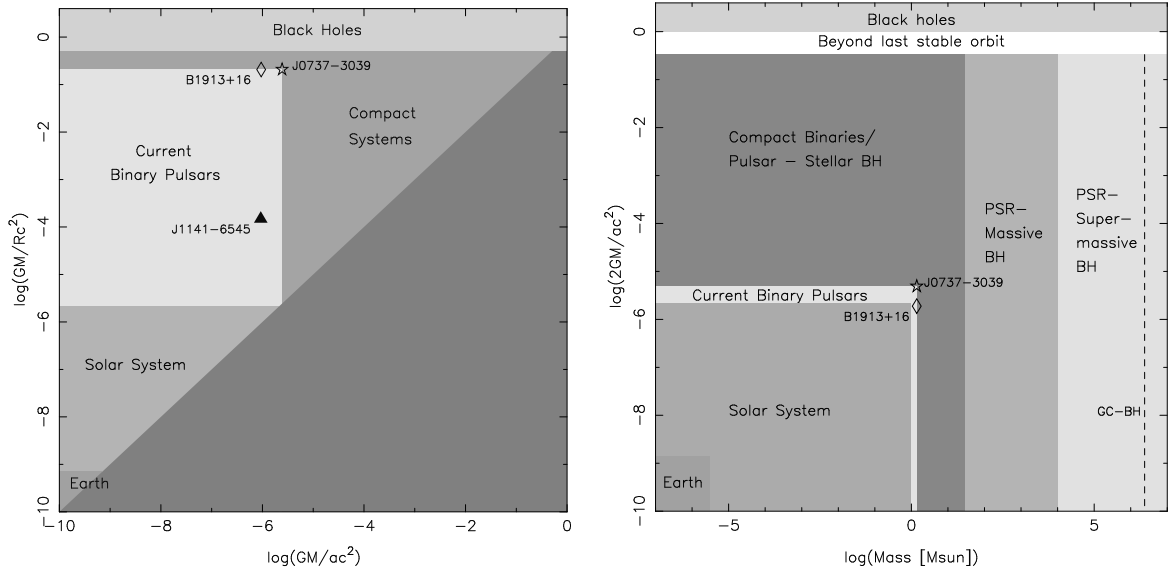

\centerline{
\begin{tabular}{cc}
\psfig{file=atest.ps,width=7.5cm} & 
\psfig{file=atest2.ps,width=7.5cm} 
\end{tabular}
}

\vspace{-0.7cm}

\caption{\label{fig:parspace} Parameter space of gravitational physics
to be probed with pulsars and black holes.  {\em (Left:)} Some
theories of gravity predict effects that depend on the compactness of
the gravitating body which is shown here (y-axis) as a function of
orbital size and probed gravitational potential  (x-axis).
Note that the lower right half of the diagram is excluded as
it implies an orbit smaller than the size of the body.
{\em (Right:)} Orbital size in units of Schwarzschild radius
as a function of gravitational mass.}
\end{figure*}

Ultra-sensitive pulsar observations with the SKA will open a new era
in fundamental probing of gravitational
physics. Figure~\ref{fig:parspace} illustrates the as yet unexplored
regions of the parameter space that can be filled with observations
made with the SKA. Clearly, one should aim for discovering pulsars in
compact orbits around a massive companion, i.e.~a stellar or even
massive black hole (BH), in order to probe the ultra-strong field
limit.  We discuss the potential for discovering such a pulsar-BH
system in the next section. We will show that a pulsar orbiting a BH
in a close orbit would be a high-precision laboratory not only for GR
in general, but for BH physics in particular \cite{pt79}.  Being timed
with the SKA, a PSR-BH system would be an amazing probe of
relativistic gravity with a discriminating power that surpasses all of
its present and foreseeable competitors \cite{de98}.

Previous pulsar tests have shown that GR's prediction for
gravitational quadrupole radiation is correct to within the
measurement uncertainties, which are currently better than 1\%
\cite{wt03}. This particular test is performed by measuring the
orbital decay of a pulsar orbit due to the emission of gravitational
waves. Pulsars can also be used to directly {\em detect} gravitational
radiation in contrast to the {\em indirect} measurements from orbital
decay in binaries. What is predicted is a stochastic background
spectrum of waves from energetic processes in the early
Universe. Pulsars discovered and timed with the SKA act effectively as
the end\-points of arms of a huge, cosmic gravitational wave detector.
This ``device'' with the SKA at its heart promises to detect such a
background, at frequencies that are below the band accessible even to
LISA (see Table~\ref{table:gw}).  We discuss the direct detection
of gravitational waves in Section~\ref{sec:level0gw}.

\begin{table}
\caption{\label{table:gw} 
Frequency range and sensitivity limit of an SKA
pulsar timing array (PTA) in comparison other
means of detecting gravitational wave emission.}
\begin{tabular}{lccc}
\hline
\hline
\noalign{\medskip}
Detector & Approximate & Sensitivity \\
\noalign{\smallskip}
         & Frequency  &  $h_0^2\Omega_{gw}$ \\
\noalign{\medskip}
\hline
\noalign{\medskip}
Advanced LIGO & $\sim100$ Hz   & $5\times 10^{-11}$\\
LISA & $\sim$ mHz & $10^{-12}$ \\     
PTA & $\sim$ nHz &  $10^{-13}$ \\
COBE & $\sim10^{-16}$ Hz & $7\times 10^{-14}$  \\
\noalign{\medskip}
\hline
\end{tabular}

{\footnotesize Non-PTA reference: Maggiore (2000)}
\end{table}


\section{Strong-field Tests of Gravity}
\label{sec:bh}

The strong-field limit of gravity is encountered when large velocities
and large gravitational potentials of massive compact objects are
involved. The effect of the latter can be estimated from the body's
gravitational self-energy, $\epsilon$. For a mass $M$ with radius $R$,
$\epsilon$ can expressed in units of its rest-mass energy,
i.e.~$\epsilon=E_{\rm grav}/Mc^2\sim -GM/Rc^2$, where $G$ is the
gravitational constant and $c$ is the speed of light. In the solar
system we find $\epsilon\sim -10^{-6}$ for the Sun and $\epsilon\sim
-10^{-10}$ for the Earth.  In contrast, for a NS, $\epsilon\sim
-0.2$, whilst for a BH $\epsilon=-0.5$.

Unlike general relativity, some alternative theories of gravity, such
as tensor-scalar theories, predict effects that depend strongly on a
body's gravitational self-energy. Such effects can be detected, for
instance, when two objects of considerable different self-energies are
moving in the same external gravitational potential or are in orbit
about each other. These phenomena, related to possible violations of
equivalence principles, can be probed in the weak-field limit of the
solar system and in the strong-field regime by observing pulsar-white
dwarf systems (e.g.~\cite{lcw+01,esp04}). Other tests can probe the
existence of preferred reference frames or violation of the
conservation of momentum \cite{sta03}.

The largest strong-field effects, however, are only encountered to
date when studying pulsars in compact binary systems.  Prime examples
are Double-Neutron Star Systems (DNS) such as the famous PSR B1913+16
system \cite{ht74} or the first-discovered double pulsar PSR
J0737$-$3039 \cite{bdp+03,lbk+04}.  As described in more detail in the
general description of the pulsar case (Cordes et al, this volume), 
pulsars enable
high-precision tests of GR by probing a number of effects such as the
possible violation of equivalence principles, conservation laws or
gravitational wave damping. However, locating the position of the
currently known binary pulsars in Figure~\ref{fig:parspace} shows that
these still do not probe those regions of the diagrams which are
populated by ultra-compact systems, in particular those with a BH
companion. These latter systems will almost certainly be discovered in
a ``Galactic Census'' of pulsars with the SKA.

Through its sensitivity, sky and frequency coverage, the SKA will
discover a very large fraction of the pulsars in the Galaxy, resulting
in about 20,000 pulsars (see Cordes et al., this
volume).  This number represents essentially all active pulsars that
are beamed toward Earth and includes the discovery of more than 1,000
millisecond pulsars (MSPs). This impressive yield effectively samples
every possible outcome of the evolution of massive binary stars,
thereby guaranteeing the discovery of systems that provide the best
opportunity for testing fundamental physics.

The computer power available when the SKA comes on-line will enable us
to do much more sophisticated acceleration searches than possible
today. At the same time, the sensitivity of the SKA allows much
shorter integration times, so that searches for compact binary pulsars
will no longer be limited. Hence, the combination of SKA sensitivity
and computing power means that the discovery rate for relativistic
binaries is certain to increase beyond the number of compact binary
systems that we can expect from a simple extrapolation of the present
numbers.  More detailed estimates come from binary evolution modelling
(e.g.~\cite{bkb02}), which imply that we should expect to find at
least 100 compact {\em relativistic} binaries, providing almost
certainly the first PSR-BH systems.

\subsection{Pulsar-Black Hole Systems}

As stars rotate, astrophysicists also expect BHs to rotate, giving
rise to both a BH spin and quadrupole moment. The resulting
gravito-magnetic field \cite{th85} gives rise to a
relativistic frame-dragging
in the BH vicinity, which causes the orbit of any test mass about the BH 
to precess if the orbit deviates from the equatorial
plane \cite{lt18}. The consequences for timing a pulsar around a BH
have been studied in detail by Wex \& Kopeikin (1999 \cite{wk99}), who
showed that the study of the orbital dynamics allows us to use the
orbiting pulsar to probe the properties of the rotating BH.

The mass of the BH can be measured with very high accuracy as done for
the DNS PSR B1913+16 \cite{wt03} or the double pulsar \cite{lbk+04}.
The spin of the BH can also be determined very precisely using the
nonlinear-in-time, secular changes in the observable quantities due to
relativistic spin-orbit coupling.  The anisotropic nature of the
quadrupole moment of the external gravitational field will produce
characteristic short-term periodicities due to classical spin-orbit
coupling, every time the pulsar gets close to the oblate BH companion
\cite{wex98,wk99}. These short-term periodicities lead to a unique
signature in the timing residuals that can be detected by regular SKA
timing.

Therefore, with SKA observations, the mass, $M$, and
both the dimensionless spin $\chi$ and quadrupole $q$,
\begin{equation}
   \chi \equiv \frac{c}{G}\; \frac{S}{M^2} \qquad \mbox{\rm and} \qquad
   q = \frac{c^4}{G^2}\; \frac{Q}{M^3} 
\end{equation}
of the BH can be determined, where $S$ is the angular momentum and $Q$
the quadrupole moment. These measured properties of a BH can be
confronted with predictions of GR.

\subsection{Cosmic Censorship Conjecture}

In the theory of GR, the curvature of space-time diverges at the
centre of a BH. The physical behaviour of this singularity is
unknown. The Cosmic Censorship Conjecture was invoked by Penrose in
1969 (see e.g.~\cite{hp70}) to resolve the fundamental concern that if
singularities could be seen from the rest of space-time, the resulting
physics may be unpredictable. The Cosmic Censorship Conjecture
proposes that singularities are always hidden within the event
horizons of BHs, so that they cannot be seen by a distant
observer.  A singularity that is found not to be hidden but ``naked''
would contradict this Cosmic Censorship. In other words, the complete
gravitational collapse of a body always results in a BH rather
than a naked singularity (e.g.~\cite{wal84}).

While the issue of whether the Cosmic Censor Conjecture is correct
remains an unresolved key issue in the theory of gravitational
collapse, we can test this conjecture by measuring the spin of a
rotating BH: In GR we expect $\chi\le1$.  If, however, SKA
observations uncover a massive, compact object with $\chi>1$ two
important conclusions may be drawn. Either we finally probe a region
where GR is wrong, or we have discovered a collapsed object where the
event horizon has vanished and where the singularity is exposed to the
outside world.  The discovered object would not be a BH as described
by GR but would represent an unacceptable naked singularity and hence
a violation of the Cosmic Censorship Conjecture \cite{he73}. In this
case, the Kerr metric would fail to be strongly asymptotically
predictable, and thus would not describe a BH \cite{wal84}.

The observational parameters that need to be determined to measure BH
spins are, at least, two Post-Keplerian parameters (see
Cordes et al., this volume, for details)
and the contributions of relativistic
spin-orbit coupling to the first and second time derivatives of the
projected semi-major axis, $\dot{x}_{\rm SO}$ and $\ddot{x}_{\rm SO}$,
and the longitude of periastron, $\dot{\omega}_{\rm SO}$ and
$\ddot{\omega}_{\rm SO}$. These contributions are likely to be small
and whilst their determination is possible with the SKA, the currently
available timing precision would almost certainly prevent such a
measurement even if a PSR-BH system were discovered today
\cite{wk99}. With the timing precision affordable with the SKA,
however, we can test the validity of the Cosmic Censorship Conjecture,
which the physical relevance of BHs depends on in large measure.

\subsection{``No-hair'' Theorem}

One may expect a complicated relationship between the spin of the BH,
$\chi$, and its quadrupole moment, $q$. However, for a rotating Kerr BH
in GR, both properties share a simple, fundamental relationship
\cite{tho80,tpm86},
\begin{equation}\label{nohair}
q = -\chi^2.
\end{equation}
This equation reflects the ``no-hair'' theorem of GR which implies
that the external gravitational field of an astrophysical (uncharged)
BH is fully determined by its mass and spin
(e.g.~\cite{st83}). Therefore, by determining $q$ and $\chi$ from
timing measurements with the SKA, we can confront this fundamental
prediction of GR for the very first time.

The secular changes caused by classical spin-orbit coupling due to the
BH quadrupole moment are typically three orders of magnitude smaller
than the changes caused by the spin-invoked relativistic spin-orbit
coupling \cite{wk99}. However, short-term periodic effects occurring
every orbit lead to a unique signature in the timing residuals that
can be used to extract the quadrupole moment.  The duration ($\sim$ closest
encounter of PSR and BH) and amplitude ($\sim$ few ns to $\mu$s) of
these periodic signatures depend on the mass of the BH and the
compactness and orientation of the orbit, all of which can be
determined accurately by regular timing observations. The detection
clearly requires the sensitivity and timing precision affordable with
the SKA and would benefit greatly from SKA multi-beaming capabilities
that allows for a very dense spacing of timing observations.

\subsection{Experimental Strategy}

About a hundred normal pulsar - stellar BH systems may be expected in
the Galactic field from population studies (e.g.~\cite{bkb02}).
However, the best prospects for studying BH properties and GR in the
strong-field limit would be given by the discovery of a PSR-BH system
with MSP companion since these pulsars provide the best timing
precision.  In standard evolutionary scenarios, however, the BH would
evolve first, hence preventing mass accretion to the pulsar and its
spin-up to small periods (e.g.~\cite{bkb02}). Nevertheless, in regions
of high stellar density, exchange interactions can lead to compact
binary systems that would otherwise not be formed (e.g.~\cite{rpr00}),
including those of MSP-BH systems. Prime survey targets would
therefore be the innermost regions of our Galaxy and Globular
Clusters. Discovering pulsars in the Galactic Centre requires high
observing frequencies of up to 15 GHz in order to combat the effects
of interstellar scattering \cite{cl97}.

Finding pulsars in orbits around massive or super-massive BHs would
allow us to apply the same techniques for determining their properties
as for the stellar counterpart \cite{wk99}.  Massive black holes
($\sim 10^{4} M_\odot$) are expected to reside in the centre of some
globular clusters, while we can also probe the super-massive BH in the
centre of our Galaxy with a mass 
$\sim 3\times 10^6 M_\odot$\cite{sog+02}. Since the spin and
quadrupole moment of a BH scale with its mass squared and mass cubed,
respectively, relativistic effects are much easier to measure for
massive and super-massive BHs. In these cases, the discovery of an
orbiting MSP is still desirable but not necessarily required.
With a complete SKA census of the Galactic pulsar population, we can
therefore probe and measure the properties of BHs on a wide range of
mass scales.


\section{Gravitational Wave Background}
\label{sec:level0gw}

The SKA will discover a dense array of MSPs distributed across the
sky. Being timed to very high precision ($<$100 ns), they act as
multiple arms of a cosmic gravitational wave (GW) detector.  Each
pulsar and the Earth can be considered as free masses whose positions
respond to changes in the space-time metric.  A passing gravitational
wave perturbs the metric and hence affects the pulse travel time and
the measured arrival time at Earth \cite{det79,fb90,rr95a}. 
If the uncertainty in the pulse arrival time is $\sigma$ and the
total observing time is $T$, then the ``detector'' is sensitive
to a dimensionless strain (or metric perturbation) of
\begin{equation}
h_c(f)\simeq \frac{\sigma}{T\sqrt{N_f}} \propto
\frac{\sigma_f}{T^{1.5}}
\end{equation}
where $h_c(f)$ describes the characteristic amplitude of the GW
background per unit logarithmic interval of frequency, $\sigma_f$ is
the square-root of the power spectral density $h(f)$, and $N_f$ the
number of normal points per period of a GW at frequency $f$.  With
observing times of a few years, pulsars are sensitive to GWs
frequencies of $f>1/T$, hence in the $\sim$nHz range.  The effect of
GWs as a pseudo-perturbation of the index of refraction in ordinary
3-space can be viewed as the independent modification of the position
of pulsars. At the same time, the other endpoint of the ``detector
arm'', the Earth, is modified in a manner that is the same for all
sources being timed, leaving a signal in the correlated timing
residuals of all pulsars which form a so-called ``Pulsar Timing
Array'' (PTA).

\subsection{Pulsar Timing Array (PTA)}

The measurement precision and accuracy of the pulsar clock is not
sufficient to detect the gravitational radiation of stellar-mass
binaries by means of a PTA. In contrast, super-massive black hole
binaries in nearby galaxies with orbital periods of a few years would
produce periodicities in pulsar arrival times of the order of 10 ns to
1 $\mu$s, which may be detectable with SKA timing
\cite{rr95a,lb01b}. Whilst the short lifetime of such massive binaries
reduces the chances of observing such systems, the SKA can
nevertheless detect the signal of a stochastic background of GW
emission produced by a large number of unresolved independent and
uncorrelated events.

A stochastic gravitational wave background should arise from a variety of
sources. Cosmological sources include inflation, string cosmology,
cosmic strings and phase transitions (e.g.~\cite{bs96,mag00}).
We can write the intensity of this GW background as
\begin{equation}
\Omega_{\rm gw}(f)=\frac{1}{\rho_c}\frac{d\rho_{gw}}{d\log{f}}
\end{equation}
where $\rho_{\rm gw}$ is the energy density of the stochastic background
and $\rho_c$ is the present value of the critical energy density for
closure of the Universe,
\begin{equation}
\rho_c=\frac{3H_0^2}{8\pi G}
\end{equation}
with $H_0\equiv h_0\times 100$ km s$^{-1}$ Mpc$^{-1}$ as the
Hubble constant (e.g.~\cite{lb01,lom02}).

\begin{figure}

\centerline{\psfig{file=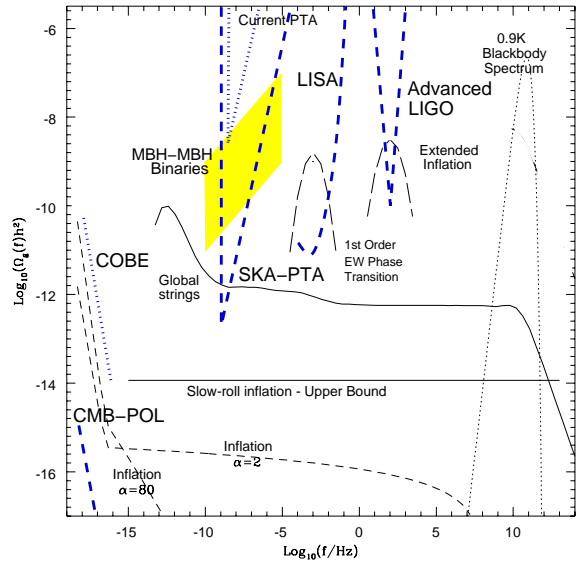,height=7.5cm}}

\caption{\label{fig:gwspec}
Summary of the potential cosmological sources of a stochastic
gravitational background, including inflationary models,
first-oder phase transitions and cosmic strings and
a primordial 0.9K black-body graviton spectrum
as presented  Battye \& Shellard (1996). We also overlay 
bounds from COBE, from current millisecond pulsar timing
and the goals from CMB polarization, LISA and Advanced
LIGO (see e.g.~Maggiore 2000). The PTA provided by
the SKA will improve on the current MSP limit by about
four orders of magnitudes. The gray area indicates
the spectrum of an additional astrophysical background 
caused by the merger of massive black holes (MBHs) 
in early galaxy formation 
(e.g.~Rajagopal \& Romani 1995, Jaffe \& Backer 2003).
For this background, $\Omega_{gw}\propto f^{2/3}$, whilst
its amplitude depends on the MBH mass function
and merger rate. The uncertainty is 
indicated by the size of the shaded area.}
\end{figure}

The frequency range covered by the PTA ($\sim$nHz) complements the
much higher frequencies accessible to Advanced LIGO ($\sim$100Hz) and
LISA ($\sim$mHz), and the extremely low frequencies probed by studies
of the Cosmic Microwave Background and its polarization (see
Table~\ref{table:gw}). For the cosmological sources a spectrum
$h_0^2\Omega_{\rm gw}(f)\sim$const.~is usually expected (see
Fig.~\ref{fig:gwspec}). With $h_0^2\Omega_{\rm gw}\propto h_c^2 f^2$,
this translates into a characteristic strain spectrum of
$h_c(f)\propto f^{-1}$.  The amplitude of the GW background from
inflation can already be constrained by COBE measurements with a safe
upper bound of $h_0^2\Omega{\rm gw}^{\rm inflation}\lesssim 10^{-14}$.
This signal appears to be too low to be observable with any GW
detector \cite{mag00}.

In contrast, a much stronger GW signal is produced in string cosmology
by the amplification of vacuum fluctuations.  Cosmic strings are
topological defects existing in grand unified theories. If they
vibrate, they create a GW background that lies in the sensitivity
range of both LISA and the PTA.  Local defects would convert larger
fractions of energy into GWs, potentially producing stronger signals
that would be even easier to detect.  However, the limits provided by
current MSP timing already exclude the existence of local strings to a
significant level \cite{lom01}.  The best current MSP timing limit is
based on a single pulsar, PSR B1855+09 \cite{lb01},
\begin{equation}
h_0^2\Omega{\rm gw}^{\rm string}<4\times 10^{-9}
\end{equation}
improving over the previously best limit \cite{ktr94} by more
than one order of magnitude.

In addition to this truly cosmological stochastic GW background, a
contribution is also expected from astrophysical processes, in
particular from the coalescence of massive BH binaries during early
galaxy evolution \cite{rr95a,jb03}. While LISA may detect individual
mergers of the more massive BH binaries, these sources also form a
stochastic background. The spectrum of this ``foreground''
contribution follows $h_c(f)\propto f^{-2/3}$ or $h_0^2\Omega_{\rm
gw}(f)\propto f^{+2/3}$ \cite{phi01}.  Whilst the exact amplitude of
this signal depends on the mass function of the massive BHs and their
merger rate \cite{jb03} (see shaded area in Fig.~\ref{fig:gwspec}),
massive BH binaries are the among primary sources expected for the
background in the frequency range detectable with LISA and the PTA
\cite{jb03,wl03,eins04}: the nHz frequency background is thought to be
dominated by BH binaries at redshifts $z\lesssim 2$ whilst more than
half of the massive BHs detectable at mHz are likely to originate at
redshifts $z \gtrsim7$.  A complementary detection of this background
and a measurement of its amplitude from nHz to mHz with LISA and the
PTA would allow to discriminate between this foreground signal from
the cosmological sources and would provide unique information about
the physics and history of BH growth in galaxies.

\nocite{bs96,mag00,rr95a,jb03}

\subsection{Experimental Strategy}

The sensitivity of the PTA scales according to
\begin{equation}
h_0^2\Omega_{\rm gw}\propto {\rm RMS}^2 f^4
\end{equation}
producing a wedge-like sensitivity curve as shown in
Fig.~\ref{fig:gwspec} (e.g.~\cite{jb03}). For timing precision that is
only limited by radiometer noise, the RMS is expected to scale with
the collecting area of the observing telescope.  In reality, the
precision is also affected by pulse phase jitter, propagation effects
in the interstellar medium and gain and polarization calibration (see
pulsar science for details). While we discuss the resulting technical
requirements such as multi-frequency capabilities and polarization
purity in \S\ref{sec:requirements}, the effects of phase-jitter mean
that not all MSPs will be suitable to be included in the PTA. Their
precision and the application of correction schemes will need to be
determined on a case by case basis. However, extrapolating from the
experience with the best performing MSPs today, we can expect the SKA
to improve on the current limit on $h_0^2\Omega_{\rm gw}$ by a factor
$10^2-10^3$!  We gain an additional factor $1/\sqrt{N_{\rm PSR}}$ from
a correlation of many pulsars, where $N_{\rm PSR}$ is the number of
timed pulsars in the PTA.  While a minimum of seven pulsars is
required for calibration and cross-correlation purposes
\cite{fb90}, ideally, a much larger $N_{\rm PSR}$ should be included
and a very large fraction of the sky should be covered by the PTA in
order to identify large-scale spatial correlations.
In summary, with the SKA sensitivity and a large number of MSPs to be
discovered and timed in the PTA, the SKA will provide a huge leap in
sensitivity of many orders of magnitude.


\section{Technical Requirements}

\label{sec:requirements}

The key observations required to achieve the outlined science
goals are 
\begin{itemize}
\item a complete Galactic census
of pulsars 
\item an ultra-deep census of Galactic
globular clusters
\item a sensitive search for pulsars in the Galactic Centre region
at high frequencies ($\sim 15$ GHz)
\item intensive observations of all PSR-BH candidates in the Galactic
field and in particular in globular clusters
\item intensive observation of all pulsars discovered
in the Galactic Centre
\item VLBI observations of all PSR-BH candidates to determine
precise distances
\item frequent ($\lesssim$ weekly) observations of a large
number of MSPs to determine the best performing clocks to
form the PTA
\item continuous timing observations of all PTA pulsars
\end{itemize}
whereas it is essential that all non-search (i.e.~timing) observations 
are performed (quasi-) simultaneously over wide range of frequencies
in order to achieve best precision by removing effects of the
frequency-dependent ``interstellar weather''.

It is important to emphasise that the technical requirements for
search and timing observations are significantly different and that
{\em both} modes of observations have to be enabled to guarantee the
success of this key science programme. The technical implications are
described in detail in the general description of the pulsar 
case (Cordes et al., this volume) and are
summarized here for completeness. The requirements are
\begin{description}
\item[{\em Configuration:}] a dense core with a significant fraction
of the total SKA area to be used in blind surveys; about 10\% of the
SKA's collecting area should be distributed over
trans-/intercontinental baselines to achieve 1 mas resolution at 5
GHz.
\item[{\em Field-of-View:}] large FOV of {\em at least} 1-deg$^2$ at
1.4 GHz that needs to be fully sampled for the blind survey and with
$\sim 50- 100$ beams/deg$^2$ for timing. The result should be
large instantaneous sky coverage with
a large number of independently steerable beams using full SKA
sensitivity.
\item[{\em Frequency range:}] frequency coverage from 500 MHz to 15
GHz to detect weak, steep-spectrum sources but also pulsars at the
Galactic Centre which are invisible due to scattering below frequencies
of 9-15 GHz. Coverage should be simultaneous to allow for simultaneous
multi-frequency observations.
\item[{\em Bandwidth:}] wide bandwidth for large sensitivity with 
adequate channelization for de-dispersion.
\item[{\em Time-resolution:}] fast temporal sampling ($<1\mu$s) for
high precision timing, less fast sampling ($\sim 50\mu$s) is required
for search observations and sampling of the full FOV.
\item[{\em Correlator:}] flexibility of the correlator/beam-former to
provide required sampling.
\item[{\em Polarization:}] full-Stokes capability with net purity of
$-40$dB to enable high precision timing.
\end{description}


\section{Conclusions}

The SKA will make some significant experimental contributions to the
quest of developing quantum gravity. It will enable us to precisely
test gravity on macroscopic scales in regions of a parameter space
that are not accessible by any other means.  It also allows us to
directly search for a gravitational wave background at a sensitivity
limit that is several orders of magnitude better than previous limits
currently achievable with pulsar timing or gravitational wave
detectors.  This is possible since a Galactic Census of pulsars with
the SKA will unlock a large number of exotic pulsar binary systems and
will produce a dense array of MSPs.  Being timed to very high
precision, the MSPs act as multiple arms of a cosmic GW detector. This
``device'', with the SKA at its heart, will be sensitive to GWs at
frequencies of nHz. Thereby complementing the much higher frequencies
accessible to Advanced LIGO ($\sim$100Hz) and LISA ($\sim$mHz), the
SKA is crucial in answering the question about the existence, nature
and composition of a GW background expected from a variety of sources,
such as coalescence of massive black hole binaries during galaxy
evolution and the evolution and decay of cosmic strings as predicted
in grand unified theories.

High precision timing observations of pulsars orbiting a stellar or
(super-)massive BHs -- only possible in the radio band and with the
SKA -- will provide unprecedented probes of relativistic gravity with
a discriminating power that surpasses all its present and foreseeable
competitors. The experiments will provide extreme limits on the most
general deviations from GR to a level a thousand times tighter than
present solar-system limits and at least an order of magnitude better
than expected from any future satellite mission. Most importantly, SKA
observations will finally address the fundamental question of whether
GR can describe nature in the ultra-strong field limit, in particular
BHs. For a wide range of BH masses, one can determine their mass, spin
and quadrupole moment to test their description in Einstein's theory
and to confront the Cosmic Censorship Conjecture and the
No-hair-theorem for the first time -- obviously a major achievement in
the history of physics!


\end{document}